\documentclass[conference]{IEEEtran}
\IEEEoverridecommandlockouts
\usepackage{cite}
\usepackage{amsmath,amssymb,amsfonts}
\usepackage{algorithmic}
\usepackage{graphicx}
\usepackage{textcomp}
\usepackage{caption}
\usepackage{subcaption}
\usepackage{tabularx}
\usepackage{makecell}
\usepackage{xcolor}
\usepackage{hyperref}
\usepackage{booktabs}
\usepackage{balance}
\def\BibTeX{{\rm B\kern-.05em{\sc i\kern-.025em b}\kern-.08em
    T\kern-.1667em\lower.7ex\hbox{E}\kern-.125emX}}
    
\usepackage{xurl} % Allows URLs to break at any character
\usepackage{hyperref}

\begin{document}

\title{AI Prototyper: A Figma Plugin for Decomposition-Based GUI Prototyping with LLMs}

\author{\IEEEauthorblockN{Tawatchai Salangsingha, Ashkan Sami, Md Zia Ullah, and Iain McGregor}
\IEEEauthorblockA{\textit{School of Computing, Engineering \& Built Environments } \\
\textit{Edinburgh Napier University }\\
Edinburgh, United Kingdom \\
\{tawatchai.salangsingha, a.sami, m.ullah, i.mcgregor\}@napier.ac.uk}
}
\maketitle

\begin{abstract}
Graphical user interface (GUI) prototyping remains a time-consuming activity that demands both design expertise and considerable manual effort. As GUI prototypes are non-code artifacts that evolve alongside requirements throughout the development cycle, automating their generation is directly relevant to software maintenance and evolution. We present AI Prototyper, an open-source Figma plugin that automates GUI prototyping through a decomposition and retrieval-augmented generation (RAG) pipeline. Given a natural-language description of a desired screen, such as a login page or a product detail card, the plugin decomposes the request into discrete GUI features, retrieves matching components from a custom 32-primitive library, and renders each component as a fully editable Figma layer with auto-layout. The pipeline uses Gemini 2.5 Flash as its LLM back-end and a Node.js/Express service. Unlike existing decomposition-based tools, AI Prototyper introduces a human-in-the-loop editing step that lets users review, modify, or extend the generated feature list before rendering, uses a different technology stack and LLM family, and supports multilingual input, producing correctly labelled interfaces in Thai, English, and Mandarin Chinese. In a preliminary evaluation, participants using AI Prototyper completed more prototypes in a fixed time window than those working manually, and expert practitioners rated the AI-generated prototypes higher across nine quality dimensions. A demonstration video is available at \url{https://youtu.be/pRoFAH7MQaE}. The source code and component library are available at \url{https://github.com/tongsalangsingha/AI-prototyper-tool}

\end{abstract}


\section{Introduction}
GUI prototyping is one of the most time-consuming activities of software development. Traditional prototyping workflows involve manual layout construction, iterative refinement, and repeated validation with stakeholders~\cite{b1, b2}. Even at the wireframe level, building a prototype that covers the expected set of interface features takes considerable effort, particularly when the screen involves standard but numerous elements such as form inputs, navigation structures, and content layouts~\cite{b3, b4}. This bottleneck means that the development cycle frequently experiences delays during the prototype phase. Since prototypes require updating as requirements evolve, tools that minimise the production costs are inherently relevant to the domain of software maintenance and evolution. For instance, when a client requests adding a search feature to an existing product listing screen, a designer can regenerate the updated prototype from a revised description in seconds rather than manually reworking the layout.

Recent work has explored using large language models (LLMs) to automate parts of this process~\cite{b5, b6}. However, single-prompt GUI generation has clear limitations: systems often miss critical components, produce inconsistent layouts, or require full regeneration to accommodate small changes~\cite{b4}. Decomposition-based approaches effectively address these limitations by segmenting a high-level design request into smaller, clearly defined sub-features, which the model processes in a sequential manner ~\cite{b7}. GUIDE~\cite{b4} demonstrated this idea by combining prompt decomposition with RAG to map features onto a component library within Figma.

We present AI Prototyper, a Figma plugin that implements the decomposition and RAG workflow to automatically generate editable GUI prototypes. Beyond the core decomposition workflow, the tool introduces a human-in-the-loop editing step that lets users inspect and modify the decomposed feature list before any components are generated, and support multilingual input, in Thai, English, Mandarin Chinese, and Malayalam. While adapting the decomposition and RAG workflow from GUIDE, the tool differs in technology stack (JavaScript/Node.js rather than TypeScript/Python), LLM back-end (Gemini 2.5 Flash rather than GPT-4o), and component library (a custom 32-component set rather than MD3). We provide empirical evidence from a two-phase evaluation with expert assessment. 

\textbf{The contribution of the paper are:}
\begin{itemize}
\item The open-source Figma plugin implementing decomposition and RAG for GUI prototyping, with human-in-the-loop editing and multilingual support.
\item The preliminary empirical evidence on productivity and prototype quality from a two-phase evaluation with expert assessment.
\end{itemize}

% The remainder of this paper is structured as follows.    \textbf{Section}~\ref{sec:aiprototyper} presents the tool's overview and architecture. \textbf{Section}~\ref{sec:evaluation} describes a preliminary evaluation and \textbf{Section}~\ref{sec:related-work} reviews related work. Finally, \textbf{Section}~\ref{sec:conclude} concludes the paper with future development.

\begin{figure*}[tb]
\centerline{\includegraphics[width=0.9\textwidth]{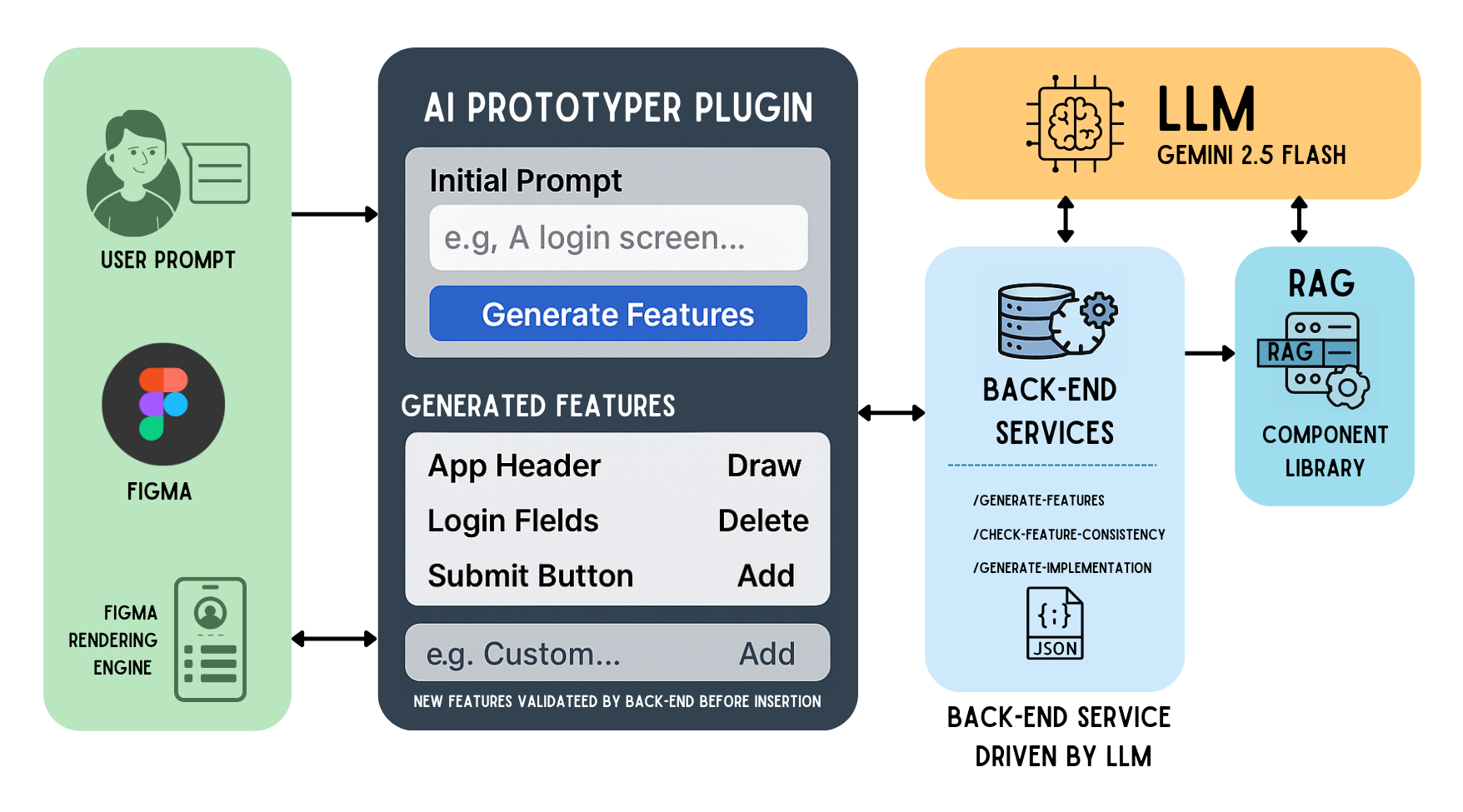}}
\caption{Overview of the AI Prototyper Architecture}
\label{fig:1}
\end{figure*}

\section{AI Prototyper}
\label{sec:aiprototyper}
\subsection{Usage Scenario}
AI Prototyper supports rapid GUI prototyping through a simple workflow.  The user opens the AI Prototyper plugin in Figma and enters a natural-language description of the desired screen. The plugin returns a decomposed feature list. The user reviews, edits, or extends the list, then triggers rendering. The output is a fully editable Figma prototype.

\subsection{Pipeline Architecture}
AI Prototyper implements a four-stage pipeline, illustrated in Figure~\ref{fig:1}.

\begin{figure}[htbp]
    \centering
    \begin{subfigure}[b]{0.48\columnwidth}
        \centering
        \includegraphics[width=\textwidth]{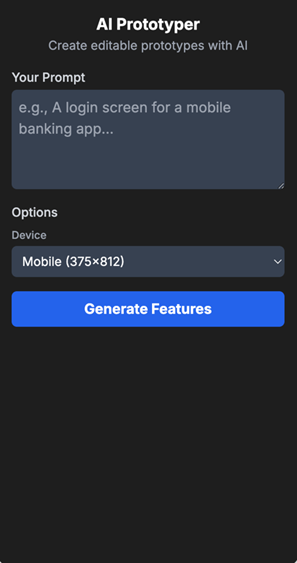}
        \caption{}
    \end{subfigure}
    \hfill % This adds horizontal space between the two images
    \begin{subfigure}[b]{0.48\columnwidth}
        \centering
        \includegraphics[width=\textwidth]{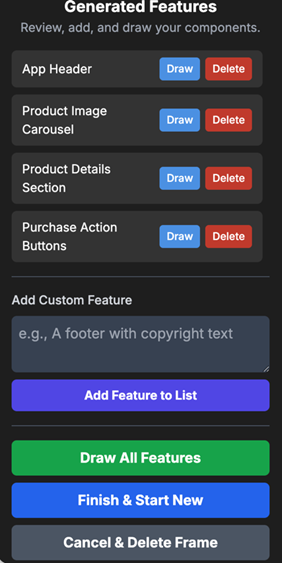}
        \caption{}
    \end{subfigure}
    \caption{AI Prototyper plug-in interface. (a) Initial Prompt View, where users supply a natural-language description and device constraints. (b) Features View showing the decomposed feature list, human-in-the-loop editing actions, and custom feature validation workflow.}
    \label{fig:2}
\end{figure}

\textbf{Stage 1: Feature Decomposition.} The user's natural-language description is forwarded to Gemini 2.5 Flash with a structured system prompt that casts the model as an expert product manager. The LLM returns a JSON array of discrete GUI features, each with a name and purpose as shown in Figure ~\ref{fig:2}b. A language rule in the prompt instructs the model to detect the input language and produce all user-facing text in that same language.
 
\textbf{Stage 2: Component Retrieval.} For each feature, the back-end performs retrieval-augmented generation over the component library. A selection prompt provides a one-line-per-component summary of the catalogue and asks the model to identify a small set of relevant components. This two-stage RAG design avoids sending the full library specification in a single pass, keeping context length manageable and improving selection accuracy.
 
\textbf{Stage 3: Component Instantiation.} A separate implementation prompt provides the full property specifications of the selected components along with the available icon inventory. The model generates a JSON array of component instances that conforms to the fixed schema, which the back-end validates before returning.
 
\textbf{Stage 4: Figma Rendering.} The plugin receives the validated JASON and renders each component as native Figma layers using auto-layout. Every element is fully editable using standard Figma operations.

\subsection{Component Library}
The custom library contains 32 components organised into seven categories: Typography (4), Input \& Form (6), Actions (3), Display \& Content (5), Navigation (4), Feedback \& Status (4), and Layout \& Structure (6). Each component is defined by a JSON schema specifying required and optional properties, valid value ranges, and type constraints. We constructed the library by analysing common mobile UI patterns and selecting components that cover login screens, product cards, social profiles, and settings pages. Using a focused 32-component set rather than a larger catalogue like MD3's 70+ entries keeps the retrieval stage compact and lets us test whether the decomposition approach works with a smaller, purpose-built library.

\subsection{Prompt Engineering}
 
Three extensions to the prompting strategy distinguish AI Prototyper from prior work. First, each prompt assigns an explicit role to the model: the decomposition prompt casts it as an expert product manager; the implementation prompt casts it as an expert UI/UX designer. Second, the two-stage RAG design (selection then instantiation) keeps each prompt focused and avoids overloading the context window. Third, all prompts include a language rule that instructs the model to write user-facing text in the same language as the user's input while keeping component names and structural property values in English so the rendering engine can process them. These prompt templates are language-agnostic; we do not apply any language-specific adaptation.

\subsection{Multilingual Support}
 
AI Prototyper accepts prompts in any language. In practice, we have tested it with Thai, English, Mandarin Chinese, and Malayalam. Thai, English, and Mandarin Chinese prompts produce interfaces with correctly translated labels on the first attempt. Malayalam, a low-resource language for current LLMs, yields structurally valid prototypes but with labels that are transliterations of English keywords rather than proper translations. The underlying layout and component hierarchy remain usable regardless of input language; only label accuracy varies with the LLM's proficiency in a given language.

\subsection{Human-in-the-Loop Editing}
 
One feature of AI Prototyper is the intermediate editing step. After decomposition, the user can review, delete, reorder, or add custom features before rendering. This acts as a lightweight checkpoint that makes the model's interpretation of the design request visible and editable, which single-pass generation tools do not provide. It allows misinterpretation to be caught before the full prototype has been generated.

\subsection{Implementation Details}
 
The plugin is built with HTML, CSS, and JavaScript, communicating with a Node.js/Express back-end via REST endpoints. Three main endpoints handle the pipeline stages: \texttt{/generate-features} for decomposition, \texttt{/check-feature-consistency} for custom feature validation, and \texttt{/generate-implementation} for component retrieval and instantiation. The back-end communicates with Gemini 2.5 Flash through Google's Generative AI SDK.
 
Both the selection and implementation prompts use a chat-history priming pattern: the system instruction is placed as the first user turn, followed by a fixed model acknowledgement, before the actual user input is sent. This pattern improves instruction adherence compared with a plain system message.
 
The Figma rendering engine maps each component type in the returned JSON to Figma API calls. Components are placed within auto-layout frames that handle alignment and spacing. The rendering engine uses string matching on component names, which is why structural property values must remain in English regardless of the prompt language.
 
The component library is stored as a set of JSON files in the back-end. Adding a new component involves creating its JSON schema, adding a one-line summary to the catalogue file for the selection prompt, and registering a rendering function in the plugin. This modular structure makes the library straightforward to extend.

\subsection{Comparison with Manual Prototyping}
AI Prototyper reduces the time required to produce a testable prototype. The decomposition step ensures that all requested features are covered, reducing the risk of missing components under time pressure. However, the tool uses a fixed component library with default styling; branded visuals, custom illustrations, and interactive navigation flows still require manual editing in Figma after generation.
 
AI Prototyper is available as open-source software,\footnote{\url{https://github.com/tongsalangsingha/AI-prototyper-tool}} with the full source code, component library, prompt templates, and installation instructions.

\section{Evaluation}
\label{sec:evaluation}
We conducted a two-phase evaluation to gather initial evidence on the tool's practical usefulness. This evaluation follows the methodology used by GUIDE~\cite{b4}, adapted to our context.

\subsection{Phase 1: Productivity}
Eleven computing undergraduates with prior Figma experience were randomly assigned to two conditions: five used AI Prototyper and six used standard Figma. Each participant completed up to four mobile prototyping tasks (login screen, product card, social profile, settings screen) during a 45-minute session. All prompts were in Thai.

\begin{table}[h]
\renewcommand{\arraystretch}{1.2}
\centering
\caption{Phase 1 Productivity}
\label{tab:productivity}
\begin{tabular*}{\columnwidth}{@{}lcccccc@{}}
% \begin{tabular}{lccccc}
\hline
Condition & N & Mean & Median & SD & \makecell{Completed/\\Possible} & Completion \\
\hline
AI Prototyper & 5 & 4.00 & 4 & 0.00 & 20/20 &100\% \\
Figma & 6 & 2.17 & 2 & 1.33 & 13/24 & 54\% \\
\hline
\end{tabular*}
\end{table}

Results are summarised in Table~\ref{tab:productivity}. All five AI-assisted participants completed every task (20/20 prototypes), while the control group completed 13 of 24 possible prototypes. A Mann--Whitney test indicated a statistically significant difference, though the small sample size means this result should be treated as preliminary ($U = 27.5$, $p = .016$, Cliff's $\delta = 0.83$).This measure captures task completion but not the interaction time spent on feature editing or post-generation refinement, which future work should address.

\subsection{Phase 2: Expert Quality Assessment}
 
Ten practitioners with professional experience in mobile or web UI development independently rated all Phase~1 prototypes on nine quality dimensions using a nine-point scale: meeting requirements, inclusion of necessary components, clarity of text, design consistency, visual design quality, information organisation, ease of interaction, minimal structural errors, and overall satisfaction. All prototypes were anonymised and evaluated regardless of completion status, so some manual-group entries may have been partial drafts.

\begin{table}[h]
\renewcommand{\arraystretch}{1.2}
\centering
\caption{Phase 2 Expert Quality Ratings}
\label{tab:quality}
\setlength{\tabcolsep}{3.5pt}
\begin{tabular}{lcccccc}
\toprule
 & \multicolumn{2}{c}{Control} & \multicolumn{2}{c}{AI Prot.} & & \\
\cmidrule(lr){2-3} \cmidrule(lr){4-5}
Dimension & Mean & Mdn & Mean & Mdn & $p$-value & $r$ \\
\midrule
A. Meets requirements         & 4.767 & 5 & 7.640 & 8 & 2.7e-18 & 0.832 \\
B. Necessary components       & 4.500 & 5 & 7.520 & 7 & 5.7e-18 & 0.824 \\
C. Clear \& appropriate text  & 4.300 & 4.5 & 7.500 & 7 & 1.1e-19 & 0.866 \\
D. Design consistency         & 4.350 & 5 & 7.400 & 7 & 1.2e-17 & 0.815 \\
E. Visual design              & 4.050 & 4.5 & 7.060 & 7 & 1.8e-17 & 0.811 \\
F. Information organisation   & 4.450 & 5 & 7.400 & 7 & 1.7e-17 & 0.812 \\
G. Ease of interaction        & 4.333 & 5 & 7.480 & 7 & 4.2e-18 & 0.827 \\
H. Minimal errors             & 4.367 & 5 & 6.720 & 7 & 1.1e-10 & 0.615 \\
I. Overall satisfaction       & 4.433 & 5 & 7.500 & 8 & 1.1e-17 & 0.817 \\
\bottomrule
\end{tabular}
\end{table}

As shown in Table~\ref{tab:quality}, AI Prototyper prototypes received higher ratings on all nine dimensions, with all differences statistically significant ($p < .001$) and effect sizes ranging from $r = 0.615$ to $r = 0.866$.

\subsection{Multilingual Observations}
 
We selected Thai as the primary language because the main study was conducted with Thai-speaking participants. This led to the multilingual observation probes exploring different LLM proficiency levels; English was included as the baseline, Mandarin Chinese represents a high-resource non-Latin script, and Malayalam a low-resource one. Each language was tested by a single native speaker as an early exploratory assessment rather than a controlled study. A full multilingual evaluation with statistical analysis  would be needed to confirm label accuracy across languages.

\begin{figure}[h]
    \centering
    \includegraphics[width=0.5\textwidth]{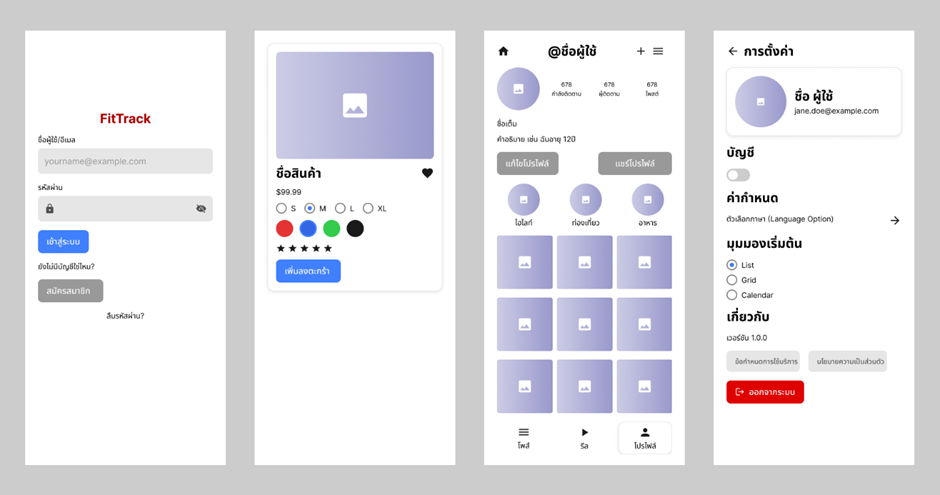}
    \caption{Example prototypes generated by AI Prototyper from Thai prompt}
    \label{fig:3}
\end{figure}

\begin{figure}[h]
    \centering
    \includegraphics[width=0.5\textwidth]{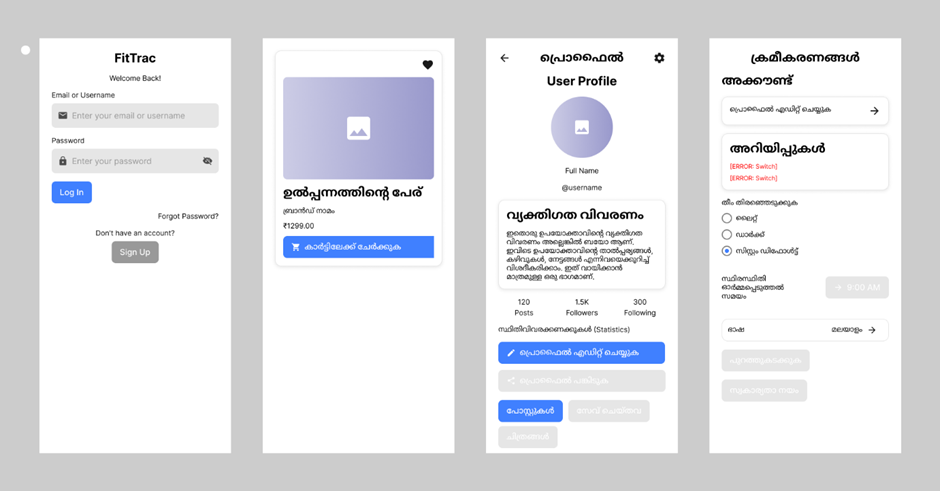}
    \caption{Example prototypes generated by AI Prototyper from Malayalam prompt}
    \label{fig:4}
\end{figure}

% \begin{figure}[h]
%     \centering
%     \includegraphics[width=0.5\textwidth]{figure_ch.png}
%     \caption{Example prototypes generated by AI Prototyper from Mandarin Chinese prompt}
%     \label{fig:5}
% \end{figure}

% \begin{figure}[h]
%     \centering
%     \includegraphics[width=0.5\textwidth]{Figure5.png}
%     \caption{Example prototypes generated by AI Prototyper from Mandarin Chinese prompt}
%     \label{fig:6}
% \end{figure}

\subsection{Limitations}
 
The sample size is small (N\,=\,11 in Phase~1, 10 expert raters in Phase~2) and all productivity participants were students rather than professional designers; a larger, more diverse sample would strengthen these findings. In Phase 1, all AI prototyper participants reached the task ceiling (4/4), confirming the tool completes tasks faster but not quantifying the productivity gain; a study with more tasks or a time-to-completion measure would address this. In Phase 2, each rater scored every prototype, so ratings are not fully independent and the p-values may be inflated; the effect sizes (r = 0.615–0.866) are more robust indicators. Moreover, some manual-group screens may have been incomplete due to time pressure, meaning the quality gap partly reflects completeness. The four tasks cover common mobile patterns but may not reflect more complex interfaces. These results should be taken as preliminary evidence supporting the tool's usefulness, not as definitive claims about its effectiveness.

\section{Related Work}
\label{sec:related-work}
GUIDE~\cite{b4} introduced decomposition-based GUI generation within Figma, combining prompt decomposition with RAG over the MD3 component library. AI Prototyper builds directly on this methodology while varying the technology stack, LLM, component library, and language scope. A direct empirical comparison between the two tools was outside the scope of this tool demonstration; both tools were evaluated against manual Figma as a shared baseline, and the results point in the same direction.
 
Several alternative paradigms exist for AI-assisted UI construction. MAxPrototyper~\cite{b8} uses a multi-agent architecture with specialised agents for styles, layouts, and component variants, but does not produce editable structures within a design tool. Misty~\cite{b9} lets developers blend aspects of reference screenshots into code through conceptual blending, targeting a different stage of the design process. The Design2Code benchmark~\cite{b10} formalised visual-to-code conversion and found that multimodal LLMs still lag behind in layout correctness. Zero-shot prompting strategies for GUI generation~\cite{b11} have shown that prompt structure affects component correctness and layout quality. LLMs have also been applied to webpage customisation from natural-language instructions~\cite{b12}, though reliability depends on DOM complexity.
 
Research on LLM structural reasoning~\cite{b6, b13, b14} supports the feasibility of schema-constrained generation, while spatial layout remains challenging for text-based models~\cite{b15}. Decomposition pipelines reduce this burden by limiting each generation step to a focused, well-defined sub-task.
 
On evaluation methodology, crowdsourcing quality varies with evaluator expertise~\cite{b16, b17, b18}. Our use of expert practitioners rather than crowdworkers follows recommendations from the HCI literature for tasks requiring consistent quality judgements.
 
Multilingual benchmarks for code generation~\cite{b19} show performance drops for non-English prompts, particularly for low-resource languages. AI Prototyper's multilingual observations are consistent with these findings.
 
\section{Conclusion and Future Work}
\label{sec:conclude}
We presented AI Prototyper, a Figma plugin that implements decomposition and retrieval-augmented generation for GUI prototyping using JavaScript/Node.js, a custom 32-component library, and Gemini 2.5 Flash. The tool takes a natural-language screen description, decomposes it into editable features, retrieves and instantiates components from the library, and renders a fully editable Figma prototype. A preliminary evaluation showed productivity gains and higher expert-rated quality compared with manual prototyping, and the pipeline produced usable outputs in four languages. 
 
Several directions remain open. The component library does not yet support style customisation such as colour themes and typography presets, which would make the tool more practical for branded projects. The pipeline currently uses a single LLM; supporting multiple models or letting users choose between them could improve output quality for different tasks. The system generates individual screens without navigation links; adding interactive page linking would move the output toward functional prototype flows. Finally, the multilingual probes could be expanded to a broader study covering more languages and multiple speakers per language.

\section{Acknowledgment}
\label{sec:acknowledgment}
The authors used Claude to improve the readability and language quality of this manuscript, and Claude Code for code development support. All content was reviewed and edited by the authors, who take full responsibility for the final work.
 
\balance
\bibliographystyle{IEEEtran}

\end{document}